\documentclass[preprint,showpacs,preprintnumbers,amsmath,amssymb,showkeys]{revtex4}

\usepackage{graphicx}
\usepackage{dcolumn}
\usepackage{bm}

\begin{document}


\title{A complex-angle rotation and \\
geometric complementarity in fermion mixing }


\author{Kee-Hwan Nam} \email{snowall@gmail.com}
\author{Kim Siyeon} \email{siyeon@cau.ac.kr}
\affiliation{Department of Physics,
        Chung-Ang University, Seoul 156-756, Korea}
\author{Seungsu Hwang} \email{seungsu@cau.ac.kr}
\affiliation{Department of Mathematics, Chung-Ang University,
Seoul 156-756, Korea}

\date{today}
\begin{abstract}
The mixing among flavors in quarks or leptons in terms of a single
rotation angle is defined such that three flavor eigenvectors are
transformed into three mass eigenvectors by a single rotation
about a common axis. We propose that a geometric complementarity
condition exists between the complex angle of quarks and that of
leptons in $\mathbb{C}^2$ space. The complementarity constraint
has its rise in quark-lepton unification and is reduced to the
correlation among $\theta_{12}, \theta_{23}, \theta_{13}$ and the
CP phase $\delta$. The CP phase turns out to have a non-trivial
dependence on all the other angles. We will show that further
precise measurements in real angles can narrow down the allowed
region of $\delta$. In comparison with other complementarity
schemes, this geometric one can avoid the problem of the
$\theta_{13}$ exception and can naturally keep the lepton basis
being independent of quark basis.
\end{abstract}

\pacs{11.30.Fs, 14.60.Pq, 14.60.St}

\keywords{neutrino mixing, quark-lepton complementarity }

\maketitle \thispagestyle{empty}

\newcommand{\cm}{\check{m}}
\newcommand{\yuk}{\mathcal{Y}}
\newcommand{\mb}{\mathbf}
\newcommand{\mc}{$\mathbf{C}~$}
\newcommand{\mh}{$\mathbf{H}~$}


\section{Introduction}

With a number of successful neutrino oscillation experiments, the
information on fermion masses and on the transformation between
the mass basis and the weak interaction basis is getting more
balanced between the quark part and the lepton part. The
Cabibbo-Kobayashi-Maskawa(CKM) matrix of quark mixing is just a
few steps from the completion. The allowed ranges of the
magnitudes of the CKM elements are narrow, $|V_{CKM}| =$
\begin{eqnarray}
    \left(
       \footnotesize \begin{array}{ccc}
        0.97383^{+0.00024}_{-0.00023} & 0.2272^{+0.0010}_{-0.0010}
            & (3.96^{+0.09}_{-0.09}) \times 10^{-3} \\
        0.2271^{+0.0010}_{-0.0010} & 0.97296^{+0.00024}_{-0.00024}
            & (42.21^{+0.10}_{-0.80}) \times 10^{-3} \\
        (8.14^{+0.32}_{-0.64}) \times 10^{-3}
            & (41.61^{+0.12}_{-0.78}) \times 10^{-3}
            & 0.999100^{+0.000034}_{-0.000004}
        \end{array}\right),
    \nonumber
    \end{eqnarray}
while the ranges of the magnitudes of
Potecorvo-Maki-Nakagawa-maskawa(PMNS) elements are still broad;
\begin{eqnarray}
    |U_{PMNS}| = \left(
        \begin{array}{ccc}
        0.79-0.86 & 0.50-0.61 & 0-0.20 \\
        0.25-0.53 & 0.47-0.73 & 0.56-0.79 \\
        0.21-0.51 & 0.42-0.69 & 0.61-0.83
        \end{array}\right)
    \label{pmnssize}
\end{eqnarray}
at the $3\sigma$ level \cite{Yao:2006px}. The unitary
transformations are conventionally described as Euler-type
subsequent operations of three separate rotations,
\begin{eqnarray}
    U(\theta_{23},\theta_{13}, \theta_{12}, \delta) =
    R(\theta_{23})R(\theta_{13},\delta)R(\theta_{12});
\end{eqnarray}
that is, through a rotation in the $1-2$ plane, another rotation in
the $1'-3$ plane, and a third rotation in the $2'-3'$ plane, the
mass basis is switched into the weak basis. From an analysis of the
global data, the three angles in CKM have the best-fit values
$\theta_{12}^q =0.229,\theta_{13}^q =0.004,$ and $\theta_{23}^q =
0.042$, and the angles in PMNS have the best-fit values
$\theta_{12}^l =0.588,\theta_{13}^l =0,$ and $\theta_{23}^l = 0.756$
\cite{GonzalezGarcia:2007ib, Maltoni:2004ei}.

Quark-lepton complementarity (QLC) is one of the theoretical
frameworks on which phenomenological data can be naturally connected
to the quark-lepton unification. In many works, the QLC idea was
built by the relation \cite{Minakata:2004xt}
\begin{eqnarray}
\theta^q_{ij} + \theta^l_{ij} = \frac{\pi}{4} \label{qlc}
\end{eqnarray}
with the mixing angle between i- and j- generations,
$\theta_{ij}$. Only $\theta_{12}$'s and $\theta_{23}$'s satisfy
the above relation such that the complementarity gives rise to
bi-maximal mixing.

The QLC represented by Eq. (\ref{qlc}) has a few points
unexplained so far. First, like bi-maximal mixing, the QLC is
obliged to keep the exception with small $\theta_{13}$'s of quarks
and leptons, which cannot make the sum maximal. Second, the
relation implies that the angles $\theta^q_{ij}$ and
$\theta^l_{ij}$ are in a plane. In low-energy theory where the
quark-lepton symmetry is broken, the common plane including those
two angles requires a strong generic connection between quark
bases and lepton bases in process of symmetry breaking, but there
is no supporting theory.

Here, we propose a model that accommodates small $\theta_{13}$'s and
that allows $\theta^q_{ij}$ and $\theta^l_{ij}$ to belong to
independent planes. In this attempt, the transformation from the
weak basis to the mass basis can be obtained by a single
complex-angle rotation about a properly defined axis
\cite{Strumia:2006db}. In other words, by a single rotation about an
axis, the weak eigenstates $(\nu_e, \nu_\mu, \nu_\tau)$ or $(d, s,
b)$ are switched into $(\nu_1, \nu_2, \nu_3)$ or $(d_1, d_2, d_3)$,
respectively. Thus, there are two complex angles, one corresponding
to quark mixing and the other corresponding to lepton mixing. We
introduce the complementarity by a relation of those two complex
angles such that
\begin{eqnarray}
\Theta_L^2 - \Theta_Q^2 = \left(\frac{\pi}{4}\right)^2,
\end{eqnarray}
where they are the orthogonal components of a hyperbola of radius
$\pi/4$. With such a geometric constraint, the model can protect the
complementarity from the $\theta_{13}$ exception. Furthermore, the
quark basis and the lepton basis are independent of each others as
implied by completely broken quark-lepton symmetry.

In Section II, the definition of the complex angle and the axis to
represent the transformation will be introduced. In Section III,
using the hyperbolic condition, the allowed range of Dirac
Charge-Parity violating(CP) phase will be predicted. It will be
shown that the more precise measurements in other angles in the
future can test the model itself, as well as the CP violation
testable in neutrino oscillation. Only the CKM-type matrix without
Majorana phases is considered as the PMNS matrix. An extended work
with more details, including Majorana phases and the physical
implication relevant to them, is in progress in other work. A
brief on the convention to deal with complex angles is attached.

\section{A single complex-angle rotation}

In both quarks and leptons, a unitary transformation between flavor
eigenstates $|f_\alpha\rangle$ and mass eigenstates $|e_i\rangle$
consists minimally of three angles and a CP phase:
    \begin{eqnarray}
        |f_\alpha\rangle = U(\theta_{12}, \theta_{13}, \theta_{23},
        \delta) |e_i\rangle. \label{fug}
    \end{eqnarray}
If the weak interaction basis is properly chosen such that the
mass matrix of up-type quarks and that of charged leptons are
diagonal, $|f_\alpha\rangle$ and $U$ in Eq. (\ref{fug}) represent
either down-type quarks and the CKM matrix in the quark sector or
neutrinos and the PMNS matrix in the lepton sector. Suppose, in a
representation, that
$|e_1\rangle=(1,0,0)^T,|e_2\rangle=(0,1,0)^T$, and
$|e_3\rangle=(0,0,1)^T$. Then $|f_\alpha\rangle=(U_{f_\alpha
1},U_{f_\alpha 2},U_{f_\alpha 3})^T$, for $f_\alpha =d,s,b$ or
$f_\alpha =e,\mu,\tau$. The $U_{f_\alpha i}$ is an element of the
transformation matrix in Eq. (\ref{fug}). The components of
$f_\alpha$ are denoted by $(f_1, f_2, f_3)$ if the specification
of `quark or lepton' is not necessary. In a real vector space, an
orthogonal set of three vectors can fit into another orthogonal
set of three vectors simply by rotating the original set about a
common axis, which can be found to be invariant under the
rotation. Likewise, the unitary transformation in Eq. (\ref{fug})
can be replaced by a rotation with a single complex angle.

\begin{figure}
\resizebox{40mm}{!}{\includegraphics{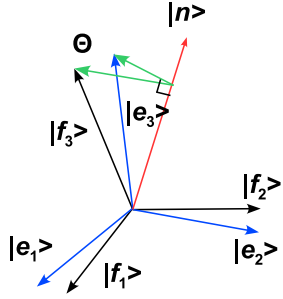}}
\caption{\label{fig:singlerot} The rotation of $\Theta$ about
$|n>$ transforms $|e_i>$'s into$|f_\alpha>$'s.}
\end{figure}

If one constructs a vector $|n\rangle$ as the axis of the rotation,
the rotation angle of a vector $|e_i\rangle$ about $|n\rangle$ is
the same as the angle between the following two vectors
$|e^\bot_i\rangle$ and $|f^\bot_\alpha\rangle$ that are orthogonal
to $|n\rangle$:
    \begin{eqnarray}
        \cos \Theta
        &=& \frac{\langle e_i^\bot|f_\alpha^\bot\rangle}
        {\sqrt{\langle e_i^\bot|e_i^\bot\rangle}
            \sqrt{\langle f_\alpha^\bot|f_\alpha^\bot\rangle}}
        \nonumber \\
        &=& \frac{\langle e_i|f_\alpha\rangle
            -\langle e_i|n\rangle\langle n|f_\alpha\rangle}
        {\sqrt{\langle e_i^\bot|e_i^\bot\rangle}
            \sqrt{\langle f_\alpha^\bot|f_\alpha^\bot\rangle}},
        \label{anglec}
    \end{eqnarray}
where
    \begin{eqnarray}
        && |f_\alpha^\bot\rangle = |f_\alpha\rangle
            - |n\rangle\langle n|f_\alpha\rangle, \nonumber \\
        && |e_i^\bot\rangle = |e_i\rangle
            - |n\rangle\langle n|e_i\rangle. \nonumber
    \end{eqnarray}
The rotation axis $|n\rangle=(n_x,n_y,n_z)^T$ has the same
components on the mass basis $|e_i\rangle$ as on the flavor basis
$|f_\alpha\rangle$, that is,
\begin{eqnarray}
    n_x|e_1\rangle + n_y|e_2\rangle + n_z|e_3\rangle
    = n_x|f_1\rangle + n_y|f_2\rangle + n_z|f_3\rangle,
            \label{axis}
\end{eqnarray}
which is a normalized vector with $|n_x|^2+|n_y|^2+|n_z|^2=1$.
Substituting Eq. (\ref{fug}) into Eq. (\ref{axis}) results in the
following combined equations:
\begin{widetext}
\begin{eqnarray}
    && c_{13}c_{12}n_x-c_{13}s_{12}n_y+s_{13}e^{-i\delta}n_z=n_x, \nonumber
    \\
        && (-c_{23}s_{12}-s_{23}s_{13}c_{12}e^{i\delta})n_x +
        (c_{23}c_{12}-s_{23}s_{13}s_{12}e^{i\delta})n_y +
        s_{23}c_{13}n_z =n_y, \\
    && (s_{23}s_{12}-c_{23}s_{13}c_{12}e^{i\delta})n_x +
    (-s_{23}c_{12}-c_{23}s_{13}s_{12}e^{i\delta})n_y +
    s_{23}c_{13}n_z =n_z. \nonumber
\end{eqnarray}
\end{widetext}
It is possible to express $|n\rangle$ immediately in terms of
mixing parameters. For instance, $|e_1^\bot\rangle,
|f_1^\bot\rangle$, the projected vectors of $|e_1\rangle,
|f_1\rangle$ on the plane perpendicular to $|n\rangle$, are,
according to Eq. (\ref{anglec}),
\begin{eqnarray}
    && |e_1^\bot\rangle = \left( \begin{array}{c}
        1-|n_x|^2 \\
        -n_x^*n_y \\
        -n_x^*n_z \\
        \end{array} \right), \\
    && |f_1^\bot\rangle = \left( \begin{array}{c}
        U_{11} \\
        U_{12} \\
        U_{13} \\
        \end{array} \right) - (n_x^*U_{11} + n_y^*U_{12} +
        n_z^*U_{13})
    \left( \begin{array}{c}
        n_x \\
        n_y \\
        n_z \\ \end{array} \right). \nonumber
\end{eqnarray}
Then, the $\cos\Theta$ in Eq. (\ref{anglec}) reduces to
\begin{eqnarray}
    \cos \Theta =
    \frac{U_{11}-n_x(n_x^*U_{11}+n_y^*U_{12}+n_z^*U_{13})}
    {\sqrt{(1-|n_x|^2)(1-|n_x^*U_{11}+n_y^*U_{12}+n_z^*U_{13}|^2})},
    \label{cotheta}
\end{eqnarray}
because $(n_x,n_y,n_z)$ is obtained in terms of mixing angles and a
phase, as is $\cos\Theta$.

The four physical parameters in the unitary mixing in Eq.
(\ref{fug}) are now all embedded in the direction of $|n\rangle$.
For a complex vector, one can remove the imaginary phase in one of
the elements by multiplying all the elements in the vector by an
overall phase factor. Together with normalization, the complex
vector $|n\rangle=(n_x,n_y,n_z)^T$ does include just four
independent parameters while the complex angle $\Theta$ does not
include any additional independent parameter. Once $|n\rangle$ is
found, then a unique $\Theta$ is determined to represent the
transformation. It can be also checked that $\cos\Theta$ is
complex unless $\sin\delta$ vanishes. As an example to find
$|n\rangle$ and $\Theta$, if one considers $\theta_{12}=0.23,
~\theta_{23}=0.042, ~\theta_{13}=0.004,$ and $\delta = 0.99$
induced from the best-fit values of the elements in the CKM
matrix, the rotation of an angle $\Theta = 0.23e^{i\varphi},~
\varphi < 0.01,$ obtained from Eq. ({\ref{cotheta}}) about the
axis
    \begin{eqnarray}
        |n_Q\rangle &=& 0.18e^{-.005i}|e_1\rangle + 0.017e^{0.868i}|e_2\rangle +
        0.983|e_3\rangle \nonumber \\
            &=& 0.18e^{-.005i}|d\rangle + 0.017e^{0.868i}|s\rangle +
        0.983|b\rangle \label{evckm}
    \end{eqnarray}
results in the same transformation as the CKM matrix does.

\section{Hyperbolic complementarity condition in $\mathbb{C}^2$ space}

As shown in the previous section, there exists a single complex
angle rotation that replaces any three-dimensional unitary
transformation. The single complex angles to replace CKM matrix in
quarks and PMNS matrix in leptons are named $\Theta_Q$ and
$\Theta_L$, respectively. They become the orthogonal components to
make a hyperbola of radius $\pi/4$ in two-dimensional complex
vector space $\mathbb{C}^2$. The geometric constraint is imposed
by
\begin{eqnarray}
    \Theta_L^2 - \Theta_Q^2 = \left(\frac{\pi}{4}\right)^2.
    \label{hyperbola}
\end{eqnarray}
$\Theta_Q$ and $\Theta_L$ are Hermitian angles as defined in Eq.
(\ref{hermitian}) \cite{Scharnhorst:1999nr} and correspond to the
absolute values of complex angles. With $\Theta_Q=0.23$ from Eq.
(\ref{evckm}), the condition is reduced to $\Theta_L=0.818$.
Knowing a Hermitian angle $\Theta_Q$ leads to the direction
$|n_Q\rangle$, and vice versa. Likewise in the Eq. (\ref{evckm})
for the CKM matrix, $|n_L\rangle$ and $\Theta_L$ can be obtained
from the parameters in the PMNS matrix.

However, taking best-fit values of the mixing angles to obtain
both $\Theta_Q$ and $\Theta_L$ does not satisfy the condition in
Eq. (\ref{hyperbola}). For the simplest example with the best-fit
values $\theta_{12}^L =0.59$ and $\theta_{23}^L = 0.76$, the
unitary transformation with $\delta^L=0$ and $\theta_{13}^L=0.18$
is equivalent to a rotation about the axis
    \begin{eqnarray}
        |n_L\rangle &=& 0.82|o_1\rangle +0.038|o_2\rangle +
        0.57|o_3\rangle \nonumber \\
            &=& 0.82|\nu_e\rangle +0.038|\nu_{\mu}\rangle +
        0.57|\nu_{\tau}\rangle
    \end{eqnarray}
by an angle $\Theta_L = 0.99$. Even though any other value of
$\theta_{13}^L$ below the upper bound 0.23 at $3\sigma$ CL is
considered, as well as any other value of $\delta^L$,
$\Theta_L=0.818$ cannot satisfy Eq. (\ref{hyperbola}) unless
values of $\theta_{12}^L$ and $\theta_{23}^L$ are far from their
current best-fit values.

\begin{figure}
{\includegraphics[width=0.23\textwidth]{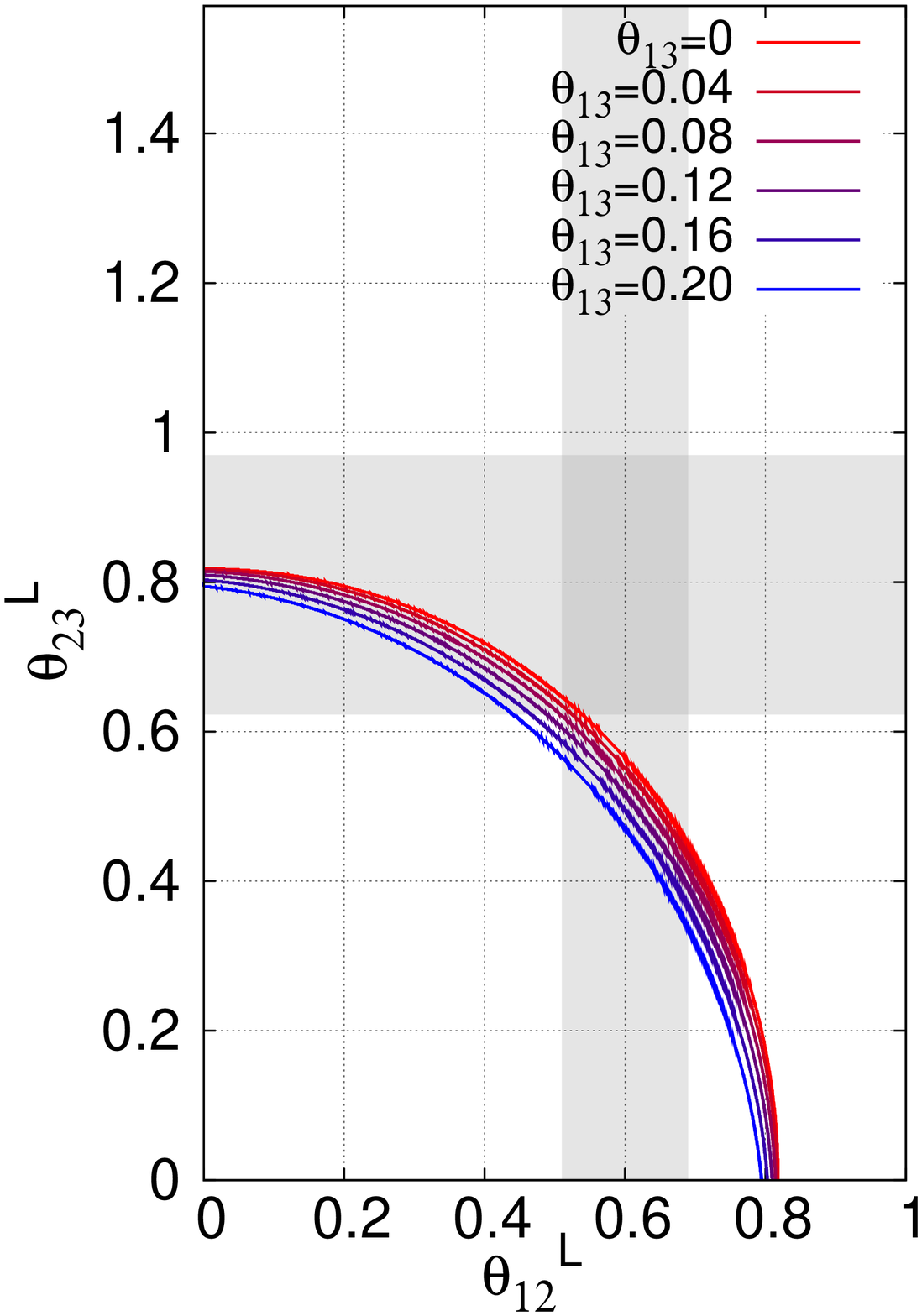}
\includegraphics[width=0.23\textwidth]{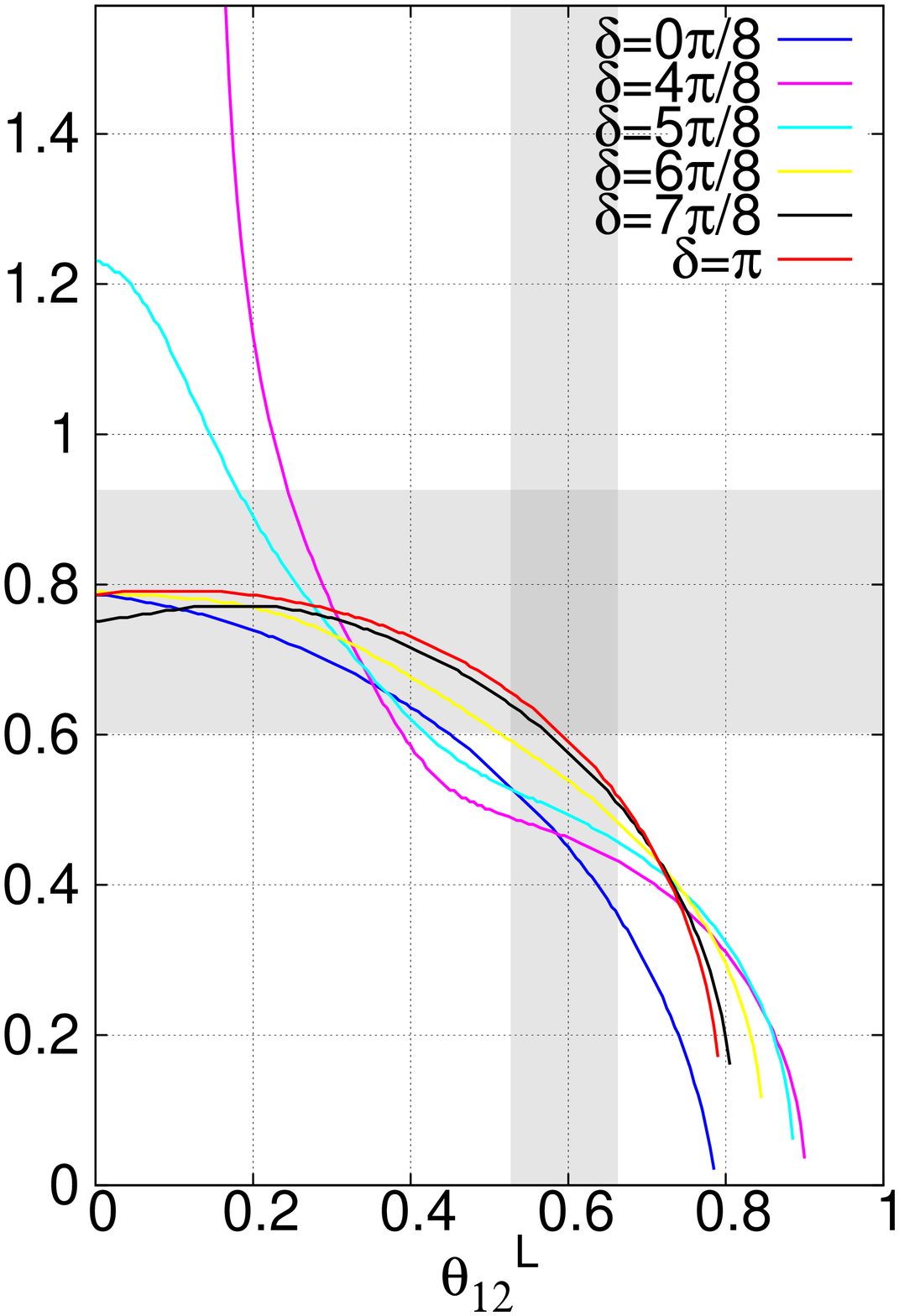}}
\caption{\label{fig:diffdelta} Plots of $\Theta_L^2 - \Theta_Q^2 =
\left(\pi/4\right)^2$ for fixed $\delta^L=0$ (left) and for fixed
$\theta_{13}=0.2$ (right). The dark bands indicate the global-fit
ranges of $\theta_{23}$ and $\theta_{12}$ at the $3\sigma$ level.}
\end{figure}

The uncertainties allowed by the current accuracy are given by the
variations, 0.62 - 0.97 for $ \theta_{23}^L$ and $0.51 - 0.69$ for
$\theta_{12}^L$ at the $3\sigma$ CL illustrated as the width of
shadow in Fig. \ref{fig:diffdelta}. The left plot in Fig.
\ref{fig:diffdelta} illustrates that the values of $\theta_{13}^L$
can satisfy the condition in Eq. (\ref{hyperbola}) if
$\delta^L=0$. Since a series of experiments on reactor neutrino
oscillations, Double Chooz, Daya Bay, and RENO, aim to determine
the value of $\theta_{13}^L$ as being larger than 0.03 in a few
years \cite{GonzalezGarcia:2007ib}, we can confirm a curve of the
constraint in $\theta_{23}^L~-~\theta_{12}^L$ space. In the case
where $\delta^L=0$, $\theta_{13}^L$ larger than 0.08 cannot be the
physical solution, as shown in the figure. However, a
$\theta_{13}^L$ whose curve does not pass the region of allowed
data in the left figure is not ruled out.

\begin{figure}
{\includegraphics[angle=270,width=0.35\textwidth]{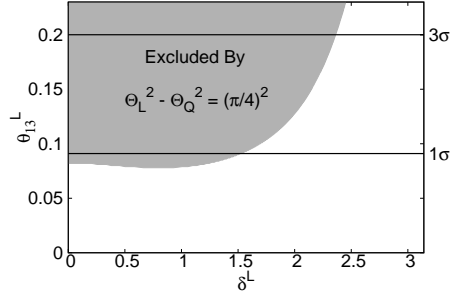}}
\caption{\label{fig:dvsq13} The allowed value of $\delta_{CP}$
with respect to $\theta_{13}$ for the ranges in $\theta_{23}$ and
$\theta_{12}$ at the $3\sigma$ CL.}
\end{figure}

The right plot in Fig. \ref{fig:diffdelta} shows that there is a
certain range in the value of $\delta^L$ with the respect to
$\theta_{13}^L=0.2$ which can make the constraint compatible with
data within $3\sigma$ CL. The allowed range in $\delta^L$ depends
on the size of $\theta_{13}^L$ as shown in Fig. \ref{fig:dvsq13},
so that the geometric complementarity condition predicts the CP
$\delta^L$ after $\theta_{13}^L$. However, the curve of
$\delta^L=\pi$ with a smaller $\theta_{13}^L$ does not approach
the central range of the data, even though $\theta_{13}^L=0$ is
the closest to the center of the allowed range, because the
deviation due to different $\delta^L$'s is determined with the
size of $\theta_{13}^L$ as the amplitude. It is possible that more
precise measurements will rule out geometric complementarity, if
they rule out $\theta_{23}^L<0.65$ and $\theta_{12}^L<0.57$.

In conclusion, complementarity and the experimental data are
compatible with each other only within a small area in neutrino
mixing angle parametric space. The constraint will predict the
value of CP $\delta^L$ when all the real angles are
better-measured. Thus, the model with its predicted CP $\delta^L$
can be tested by using long base-line oscillations like JHF in the
near future \cite{Itow:2001ee}, or by using astronomical neutrino
bursts \cite{Takahashi:2006fe}.

\appendix

\section{Complex vector space}
It is useful to introduce the concept of an angle in complex
vector spaces \cite{Scharnhorst:1999nr}. In a finite-dimensional
real (Euclidean) vector space $\mathbb{V}_\mb{R} (\simeq \mb{R}_n,
n \in \mb{N}, n\geq2)$, the angle between two vectors $\mb{A}$ and
$\mb{B}$ is defined in terms of the scalar product.
    \begin{eqnarray}
        \cos\Theta(\mb{A},\mb{B})
        =\frac{(\mb{A},\mb{B})_\mb{R}}{|\mb{A}||\mb{B}|}, \nonumber
    \end{eqnarray}
where $(\mb{A},\mb{B})_\mb{R}=\sum_{k=1}^n A_k B_k$ and
$|\mb{A}|=\sqrt{(\mb{A},\mb{A})_\mb{R}}$. There are more than one
definition of the angle between two vectors $\mb{a}$ and $\mb{b}$
in a finite-dimensional complex vector space $V_\mb{C} (\simeq
\mb{C}_n, n \in \mb{N}, n\geq2)$. A complex angle can be defined
as
    \begin{eqnarray}
        \cos\Theta_\mb{C} (\mb{a},\mb{b})
        =\frac{(\mb{a},\mb{b})_\mb{C}}{|\mb{a}||\mb{b}|}
    \end{eqnarray}
by using the Hermitian product
$(\mb{a},\mb{b})_\mb{C}=\Sigma^n_{k=1} a_k^\dagger b_k$ for any
pair of vectors $a,b \in \mathbb{V}_\mb{C}$. The cosine of the
complex angle can be rephrased, in general, as $\cos\Theta_\mb{C}
(\mb{a},\mb{b})
        =\rho e^{i\varphi}, (\rho\leq 1).$
 It is useful to introduce the definitions
of the Hermitian angle and the Euclidean angle, and their
difference. Hermitian angle $\Theta_\mb{H}$ is defined such that
    \begin{eqnarray}
        && \cos\Theta_\mb{H} (\mb{a},\mb{b}) =
        |\cos\Theta_\mb{C} (\mb{a},\mb{b})|=\rho,\label{hermitian} \\
        && 0 \leq \Theta_\mb{H} \leq \pi/2, \nonumber
    \end{eqnarray}
where $\varphi = \varphi(\mb{a},\mb{b}), (-\pi \leq \varphi \leq
\pi)$ is called the pseudo angle of  two vectors. The Euclidean
angle between two vectors in $\mathbb{V}_\mb{C}$ depends on the
vector space $\mathbb{V}_\mb{R}(\simeq \mb{R}_{2n})$ isometric to
$\mathbb{V}_\mb{C}$:
    \begin{eqnarray}
        \cos\Theta_\mb{E} (\mb{a},\mb{b}) = \cos\Theta_\mb{E}(\mb{A},\mb{B})
        =\frac{(\mb{A},\mb{B})_\mb{R}}{|\mb{A}||\mb{B}|},
    \end{eqnarray}
where the components of the vectors $\mb{A},\mb{B}$ are related
with those of $\mb{a},\mb{b}$ in the following way: $A_{2k-1}=
\mathbf{Re} a_k$ and $A_{2k}= \mathbf{Im} a_k, k=1..n.$ A simple
relation exists between the Hermitian angle $\Theta_\mb{H}$ and
the Euclidean angle $\Theta_\mb{E}$,
    \begin{eqnarray}
        \cos\Theta_\mb{E} (\mb{a},\mb{b}) =
        \cos\Theta_\mb{H} (\mb{a},\mb{b}) \cos\varphi.
    \end{eqnarray}

\begin{acknowledgments}
This research was supported by Chung-Ang University research
grants in 2005.
\end{acknowledgments}


\begin{thebibliography}{99}

\bibitem{Yao:2006px}
  W.~M.~Yao {\it et al.}  [Particle Data Group],
  J.\ Phys.\ G {\bf 33}, 1 (2006).

\bibitem{GonzalezGarcia:2007ib}
  M.~C.~Gonzalez-Garcia and M.~Maltoni,
  arXiv:0704.1800 [hep-ph], 2007.

\bibitem{Maltoni:2004ei}
  M.~Maltoni, T.~Schwetz, M.~A.~Tortola and J.~W.~F.~Valle,
  New J.\ Phys.\  {\bf 6}, 122 (2004)

\bibitem{Minakata:2004xt}
  H.~Minakata and A.~Y.~Smirnov,
  Phys.\ Rev.\  D {\bf 70}, 073009 (2004)
  H.~Minakata,
  arXiv:hep-ph/0505262, 2005.
  M.~A.~Schmidt and A.~Y.~Smirnov,
  Phys.\ Rev.\  D {\bf 74}, 113003 (2006)
  S.~K.~Kang, C.~S.~Kim and J.~Lee,
  Phys.\ Lett.\  B {\bf 619}, 129 (2005)

\bibitem{Strumia:2006db}
  A.~Strumia and F.~Vissani,
  arXiv:hep-ph/0606054, 2006.

\bibitem{Scharnhorst:1999nr}
  K.~Scharnhorst,
  Acta Appl.\ Math.\  {\bf 69}, 95 (2001)

\bibitem{Itow:2001ee}
  Y.~Itow {\it et al.}  [The T2K Collaboration],
  arXiv:hep-ex/0106019, 2001.

\bibitem{Takahashi:2006fe}
  R.~Takahashi and S.~Nagataki,
  J.\ Korean Phys.\ Soc.\  {\bf 49}, 1818 (2006).
  N.~Kawanaka, S.~Mineshige and S.~Nagataki,
  J.\ Korean Phys.\ Soc.\  {\bf 49}, 1827 (2006).
  K.~Murase and S.~Nagataki,
  J.\ Korean Phys.\ Soc.\  {\bf 49}, 1834 (2006).




\end{thebibliography}
\end{document}